\documentclass[jps,reprint,onecolumn,superscriptaddress,a4paper,nofootinbib]{revtex4-1}
\usepackage{graphicx}
\usepackage{footnote}
\usepackage{bm}
\usepackage{amsmath}
\usepackage{amsfonts}
\usepackage{microtype}
\usepackage{textcomp}
\usepackage{mathtools}
\usepackage{lineno}
\usepackage{chemfig}
\usepackage{palatino}
\usepackage{soul}
\usepackage{cancel}
\usepackage[]{subfig}
\DeclareCaptionLabelSeparator{frenchsep}{ : }
\usepackage{mwe}
\usepackage[section]{placeins}
\usepackage[font={small}]{caption}
\usepackage[pdfpagemode=UseNone,colorlinks=true,linkcolor=blue,citecolor=blue]{hyperref}

\begin{document}

\title{Tunable electronic structure and stoichiometry dependent disorder in Nanostructured VO$_x$ films.}
\author{A. D’Elia}
\affiliation{IOM-CNR, Laboratorio TASC, Basovizza SS-14, km 163.5, 34149 Trieste, Italy.}
\author{S.J. Rezvani}
\affiliation{Laboratori Nazionale di Frascati, INFN, Via Enrico Fermi 40, Frascati, Italy.}
\affiliation{IOM-CNR, Laboratorio TASC, Basovizza SS-14, km 163.5, 34149 Trieste, Italy.}
\author{N. Zema}
\affiliation{ISM-CNR, Istituto Struttura della Materia, Area della Ricerca di Tor Vergata, Via del Fosso del Cavaliere, Roma, Italy.}
\author{F. Zuccaro}
\affiliation{ISM-CNR, Istituto Struttura della Materia, LD2 Unit, Basovizza Area Science Park, 34149 Trieste, Italy.}
\author{M. Fanetti}
\affiliation{Materials Research Laboratory, University of Nova Gorica, Vipavska 13, 5000 Nova Gorica, Slovenia.}
\author{Blaž Belec}
\affiliation{Materials Research Laboratory, University of Nova Gorica, Vipavska 13, 5000 Nova Gorica, Slovenia.}
\author{B. W. Li}
\affiliation{National Synchrotron Radiation Laboratory, University of Science and Technology of China, Hefei 230029, People’s Republic of China.}
\author{C.W. Zou}
\affiliation{National Synchrotron Radiation Laboratory, University of Science and Technology of China, Hefei 230029, People’s Republic of China.}
\author{C. Spezzani}
\affiliation{Elettra-Sincrotrone Trieste, Basovizza I-34149, Italy.}
\author{M. Sacchi}
\affiliation{Synchrotron SOLEIL, L’Orme des Merisiers, Saint-Aubin, BP 48, 91192 Gif-sur-Yvette Cedex, France.}
\affiliation{Institut des NanoSciences de Paris, UPMC, CNRS UMR 7588, 4 Place Jussieu, 75005 Paris, France.}
\author{A. Marcelli}
\affiliation{Laboratori Nazionale di Frascati, INFN, Via Enrico Fermi 40, Frascati, Italy.}
\affiliation{ISM-CNR, Istituto Struttura della Materia, LD2 Unit, Basovizza Area Science Park, 34149 Trieste, Italy.}
\affiliation{Rome International Centre for Material Science Superstripes, RICMASS, Via dei Sabelli 119A, 00185 Rome, Italy.}
\author{M. Coreno}
\affiliation{ISM-CNR, Istituto Struttura della Materia, LD2 Unit, Basovizza Area Science Park, 34149 Trieste, Italy.}

\begin{abstract}
We present and discuss an original method to synthesize disordered Nanostructured (NS) VO$_x$ films with controlled stoichiometry and tunable electronic structures. In these NS films, the original lattice symmetry of the bulk vanadium oxides is broken and atoms are arranged in a highly disordered structure . The stoichiometry-dependent disorder as a function of the oxygen concentration has been characterized by in-situ X-ray Absorption Near-Edge Structure (XANES) spectroscopy identifying the spectroscopic fingerprints. Results show structural rearrangements that deviate from the octahedral symmetry with different coexisting disordered phases. The modulation of the electronic structure of the NS films based on the resulted stoichiometry and the quantum confinement in the NS particles are also discussed. We demonstrate the possibility to modulate the electronic structure of VO$_x$ NS films accessing new disordered atomic configurations with a controlled stoichiometry that provides an extraordinary opportunity to match a wide number of technological applications. 

\end{abstract}
\maketitle
\section{Introduction}
Vanadium oxides have recently been extensively studied due to their various possible stoichiometric structures characterized by different oxidation states varied from V$^{+2}$ to V$^{+5}$ as well as mixed-valence oxides (Magneli and Wadsley series) \cite{Schwingenschlogl2004} that renders an equally large variety of electronic and physico-chemical properties. For instance, two of the oxides, namely V$_2$O$_5$ and V$_6$O$_{13}$, have been deeply investigated for possible applications as cathode material in Li-ion batteries \cite{Chan2007, Xu2016, Shi2018}. Vanadium oxides also bear the seed of a strong electronic correlation. VO$_2$, V$_2$O$_3$ and the mixed-valence V$_n$O$_{2n-1}$ (n=3-6, 8 and 9) systems all exhibit first-order metal-insulator transitions (MIT) characterized by a nanoscale phase separation \cite{Marcelli2017, McLeod2017}. As a consequence accessing the nanoscale may be extremely interesting for this class of oxides.

Low dimensional materials are well known for their unique and tunable properties \cite{Jagiello20,Pasqualini2017175,SARTALE20132005} due to the modified electronic structure by quantum size effects and effective surface area \cite{Rezvani2016,Pinto2016,Rezvani20162,Pinto2018}. Hence, synthesizing nanostructured vanadium oxides with possible tunability of the electronic structure, e.g. quantum confinement enhancement of the Work Function \cite{DElia2020}, can be of great value in a vast variety of applications.

The lattice order is one of the additional parameters that can be used for modulation of the electronic parameters in nanomaterials in general and in vanadium oxides in particular, based on the desired applications \cite{Fan2014,Carturan2015116,Rezvani20162}. However, while methods to slightly modify the sample lattice, e.g. through the application of epitaxial strain \cite{Fan2014}, are widely used, the development of a technique able to produce disordered samples with controllable stoichiometry is in its early stage despite its potential.

For instance, the amorphous VO$_x$ exhibits superior intercalation properties with respect to the crystalline counterpart because due to the large number of the dangling bond within the defective structure \cite{Uchaker2014,Chae2014,Wu2018}. Accordingly, the possibility to develop a method to synthesize atomically disordered NS VO$_x$ films with large surface area and controllable stoichiometry could lead to the advancement of several technological applications. 

Supersonic Cluster Beam Deposition (SCBD) method is a well-known synthesis method able to produce cluster assembled NS materials with unique properties \cite{DElia2020,Borghi2018,Barborini2011}. This method offers considerable advantages for neutral cluster manipulation based on effusive beams making the seeded supersonic beam approach very powerful for the deposition of nanostructured films. Pulsed Micro-plasma Cluster Source (PMCS), is a cluster source characterized by high deposition rate, control over cluster growth and landing energy of the cluster beam \cite{Wegner2006,Piseri2004,Barborini1999}.

In the present work, present the synthesis NS vanadium oxides with controllable stoichiometry and disorder using the SCBD method combined with a PMCS. The coupled experimental arrangement facilitated, for the first time, in-situ characterization of disordered NS vanadium oxides films with X-ray Absorption Near Edge Structure (XANES) spectroscopy. XANES is associated with the electronic transitions from core shells to unoccupied states following a photo-absorption process, possessing the unique capability to probe concurrently the stoichiometry and local structure of the sample \cite{Bachrach1992}. The fabricated films have been characterized at distinct oxygen concentration stages, the structural evolution of the VO$_x$ as a function of oxygen content is discussed considering the effect of structural symmetry, surface area and quantum confinement within the structures. Our results improve the understanding of disordered vanadium oxide structures and encourage their exploitation in a variety of novel applications.


\section{Experimental}
The NS VO$_x$ films were deposited using a combination of the PMCS and the SCBD method.
The PMCS is a pulsed-cluster source driven by a high-power pulsed electric discharge. It was operated with a vanadium cathode (6 mm diam. rod, purity 99.9 \%, EvoChem GmbH) generating a supersonic beam of vanadium or vanadium oxide clusters. To obtain homogeneously oxidized nanoparticles, Ar (high purity Ar: 99.9995\%, SIAD) as the carrier gas was used, mixed with a controlled amount of oxygen resulting in an Ar-O$_2$ gas mixture. In the experimental layout, the supersonic cluster beam produced by the PMCS is deposited in-situ on a copper substrate, which is electrically isolated from the rest of the experimental chamber. The base pressure of the experimental chamber was $1\times10^{-9}$ mbar while operating the PMCS raised it to $\sim5\times10^{-9}$ mbar due to the carrier gas injected by the cluster source.
\begin{figure} [h!]
\includegraphics[width=0.4\textwidth]{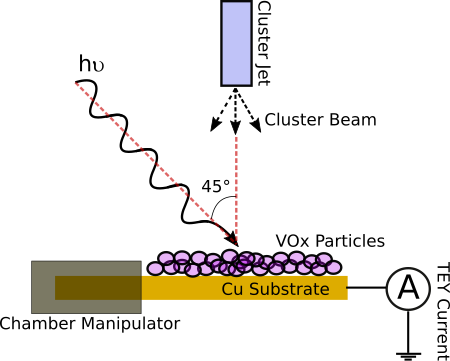}
\caption{Layout of the experimental setup including cluster beam and the TEY acquisition. A copper substrate is positioned normal to the cluster beam and with 45$^\circ$ to the incident X-ray beam. The TEY current is collected via the copper substrate.}
\label{setup}
\end{figure}
To synthesize samples with different stoichiometry, we fed the PMCS with an Ar-O$_2$ gas mixture with an O$_2$ molar percentage in the range of 0-1.25 \%. The working parameters of the PMCS have been adjusted to maximize the deposition rate. The delay between the gas injection and the discharge firing was 0.51-0.64 ms; the discharge operating voltage 0.9 kV and the discharge duration 60 $\mu$s; the pulsed-valve aperture was between 196 and 220 $\mu$s; the pulse repetition rate 3 Hz and the Ar-O$_2$ pressure feeding the Parker valve was 60 bar. All the samples have been synthesized using Ar or Ar-O$_2$ gas mixture as specified elsewhere \cite{DElia2020}. The deposited films have been studied in-situ using XANES spectroscopy.
The UHV chamber has been coupled to the PMCS, available at the TASC laboratory \cite{DeSimone2012}, and equipped with an XYZ manipulator as depicted in Figure \ref{setup}.
The XANES measurements have been performed at the Circularly Polarized beamline of Elettra synchrotron radiation facility using the CryoAlp UHV experimental chamber \cite{Marcelli2018}. Total Electron Yield (TEY) spectra have been acquired measuring the drain current, connecting the sample to ground through a Keithley 486 picoammeter (see Figure \ref{setup}). The sample was kept normal to the cluster beam while the angle between the incident photons and the sample surface was kept fixed ($\theta$=45$\pm$5$^\circ$).
Furthermore, the $ex-situ$ XANES measurements on the crystalline sample, synthesized using the Molecular Beam Epitaxy (see Ref\cite{Fan2014}), was performed using the IRMA experimental chamber \cite{Sacchi2003} and used as a reference in our experiment. All the XANES spectra have been normalized to the incident photon flux. The morphology and the size of the clusters have been analyzed by Transmission Electron Microscopy (TEM) through a field-emission TEM (JEM2100F-UHR, JEOL) operated with beam energy 200 KV.

\section{Results and discussions}
\begin{figure} [h!]
\includegraphics[width=0.5\textwidth]{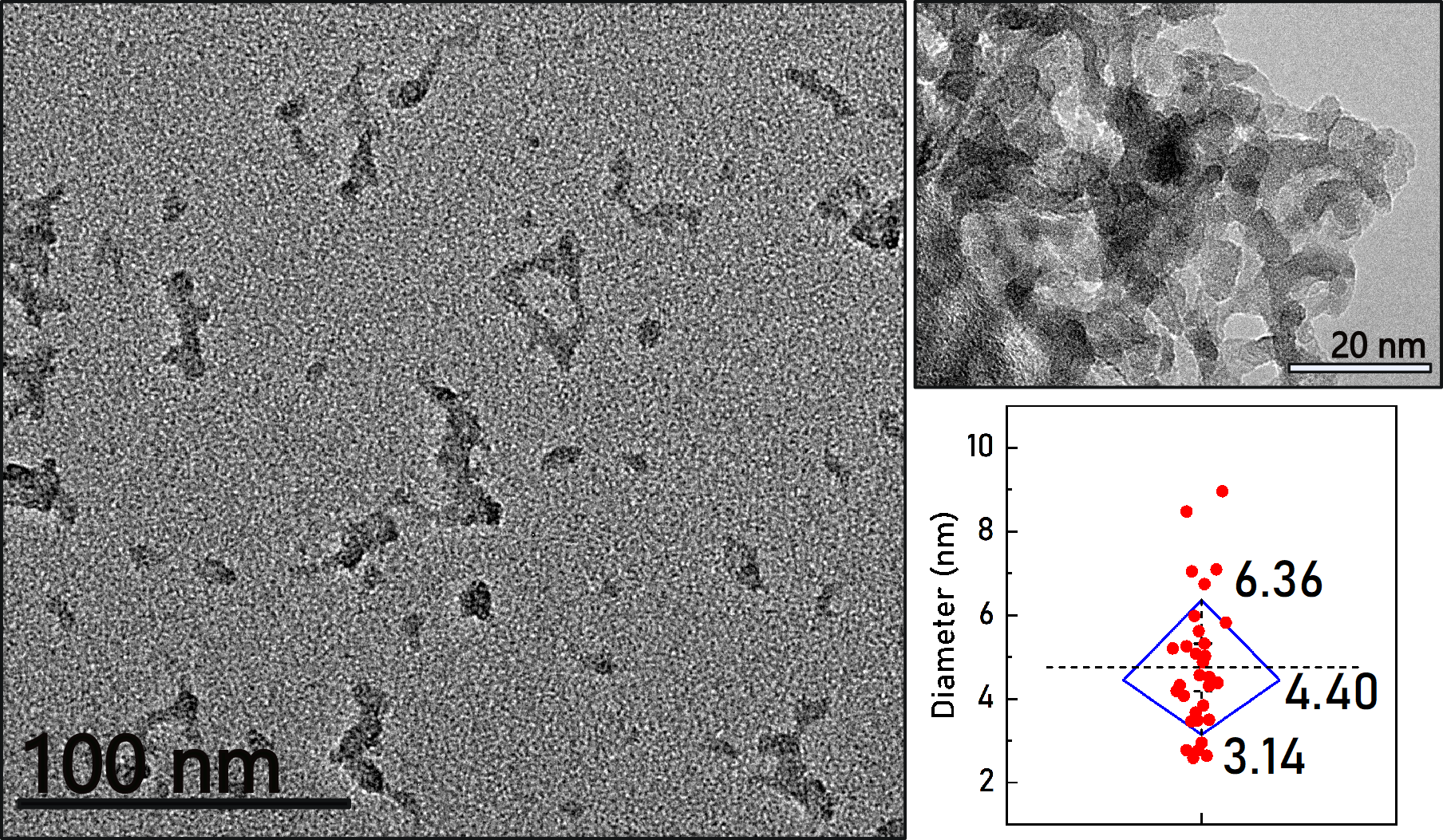}
\caption{TEM image of the nanoparticle aggregates nanostructured film for VO$_{2.5}$ along with the size distribution of the particles. The particles show a size distribution between $\sim$ 3 to $\sim$ 6 nm with an average value of d= 4.4 nm, the particles count are spread on the x axis for better visibility.}
\label{tem}
\end{figure}

\textbf{Morphology}: The morphology and size of nanoparticles deposited by SCBD have been investigated using TEM. A deposition of 1 second has been performed on a TEM grid with an amorphous C supporting film. 
The Figure \ref{tem}  represents theTEM image of  the sample with oxygen concentration corresponding to VO$_{2.5}$. As it it could be seen, the nanoparticles form agglomerates. Even though the nanoparticles aggregate into large agglomerates, the individual nanoparticles are visible in the magnified image. Despite the intense aggregation of nanoparticles that makes the size distribution assessment with precision rather difficult, the size of the nanoparticles could be measured in the range 3-6 nm with an average particle size of $\sim$ 4.4 nm.
No large differences have been observed for samples deposited with various carrier gasses and for different oxidation states. 

\begin{figure} [h!]
\includegraphics[width=0.7\textwidth]{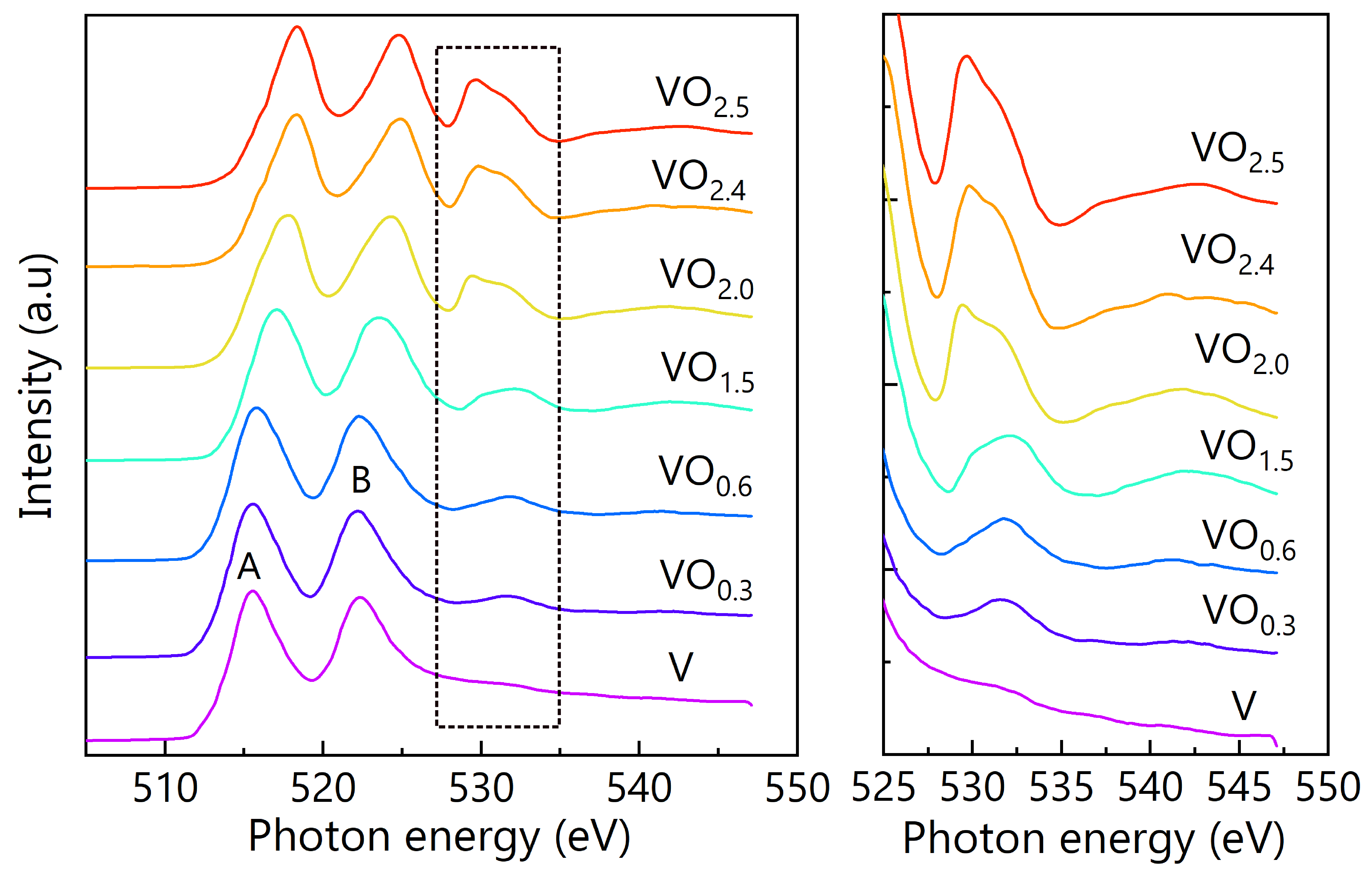}
\caption{Left: XANES spectra of different VO$_x$ films at the V L$_{2,3}$ edges and the O K-edge. Right: Comparison of the O K-edge XANES spectra of VO$_x$ samples.}
\label{VO}
\end{figure}

\textbf{Local order}: 

Different NS VOx films have been synthesized adding a small percentage of oxygen within the carrier gas as described in Ref \cite{DElia2020}. The samples have been deposited and investigated in-situ using XANES spectroscopy at V L$_{2,3}$ and O K edges. From now on to identify the different VO$_x$ samples we use the stoichiometric ratio defined as the ratio $x$= [No. O atoms/ No. V atoms], in the range $0\leq x\leq2.5$.
The XANES spectra collected across the V L$_{2,3}$ and O K edges for different VO$_x$ samples are shown in Figure \ref{VO}.
The O K-edge spectra of the VO$_x$ samples (Figure \ref{VO} right panel), reveals a significant evolution of the electronic structures in the samples with distinct stoichiometry. This evolution can be understood in terms of crystal field splitting and metal-ligand hybridization. In VO$_x$, O atoms arrange in an octahedral geometry around the vanadium atom generating a cubic crystal field \cite{Hermann2001, DeGroot1989}, which splits the degenerate $3d$ manifold into three t$_{2g}$ and two e$_g$ levels. Because of the $3d-2p$ hybridization, the energy levels are the superposition of $p$ and $d$ electrons, thus the contribution of the hybridized t$_{2g}$ and e$_g$ levels in the empty states can be observed at the O K edge XANES spectrum. As can be seen in Figure \ref{VO}, the O K edge XANES exhibit two main contributions at $\sim 530 $ eV and $\sim 532.5$ eV, whose intensity ratio overturns increasing the oxidation state.

The high energy component observed in O K-edge spectra can be associated to the e$_g^*$ bands while the low energy feature to the $t_{2g}^*$ \cite{DeGroot1989}. Since XANES probes the empty DOS, $t_{2g}^*$ and e$_g^*$ signals can be observed only when t$_{2g}$ and e$_g$ electrons are bound with the O $2p$ electrons. The presence of $t_{2g}^*$ in VO$_{0.6}$ samples (the weak component around 530 eV) can be attributed to the hybridization of V-$3d$ and O-$2p$ electrons in agreement with a previous Auger investigation, which predicts a mixed $4s-3d$ valence band for this sub-oxide \cite{DElia2020}. Increasing the oxygen content, the $3d-2p$ hybridization increases with the inversion of the intensity of the e$_g^*$ and $t_{2g}^*$ contributions for $x=1.5$, again in agreement with previously reported results \cite{Chen1994,Chen1997,DeGroot1989,Zimmermann1998}. Moreover, the e$_g^*$ and $t_{2g}^*$ features are broad and not separated, even at the highest oxidation states. This is a fingerprint of the distorted arrangement of the ligand octahedron around the metal atom \cite{Kasatikov2019} indicating that the VO$_x$ NS films are locally disordered.

\begin{figure} [h!]
\includegraphics[width=0.5\textwidth]{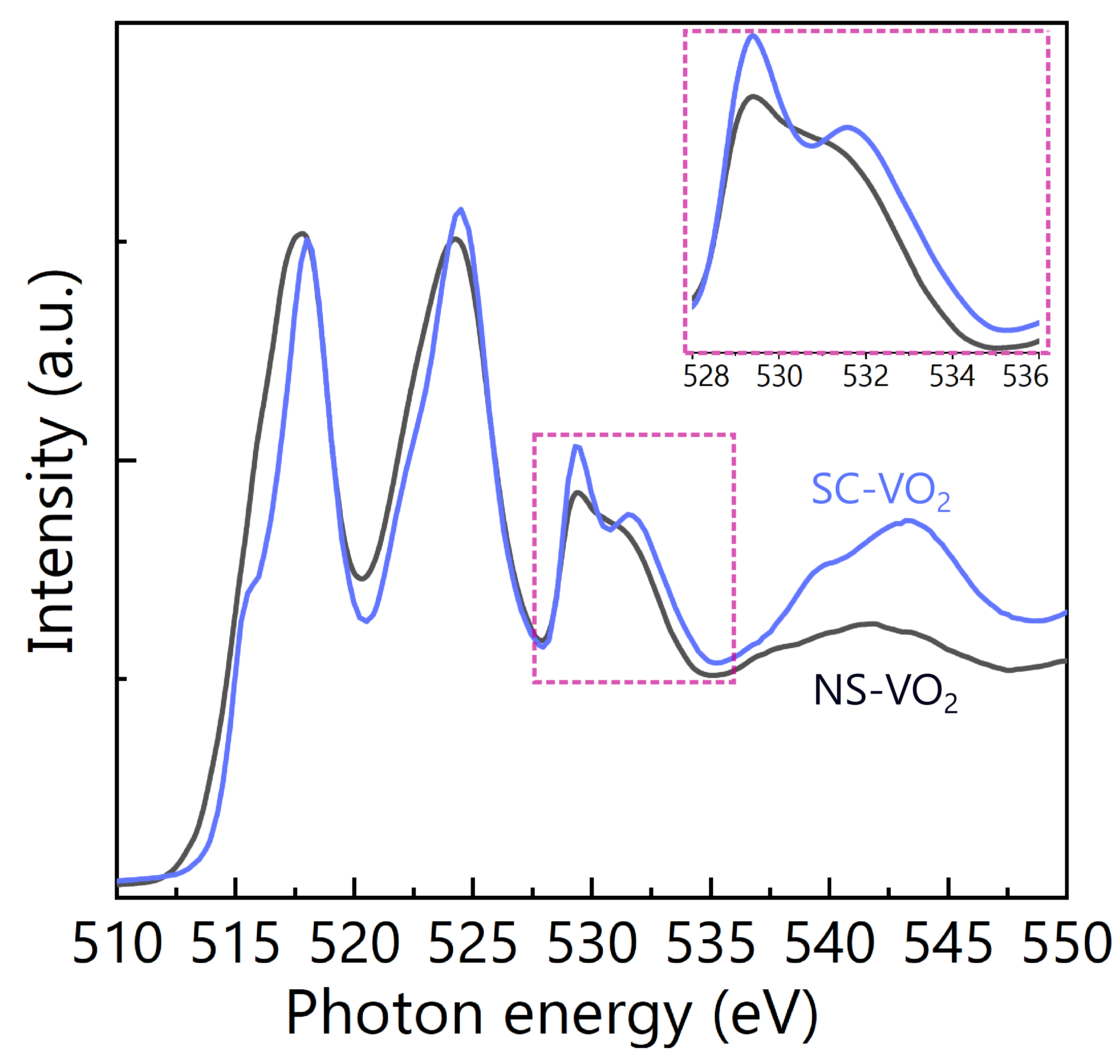}
\caption{Comparison between the O K edge of the NS  VO$_2$ film and that of a single crystalline (SC) film.}
\label{OK}
\end{figure}

The presence of eminent local disorder in VO$_x$ nanostructures can be further confirmed by comparison of the VO$_2$ NS sample XANES spectrum with that of a single crystalline (SC) VO$_2$ (see figure \ref{OK}).
In the crystalline sample spectra, the e$_g^*$ and $t_{2g}^*$ features are well separated pointing to a strong and homogeneous crystal field and therefore to a locally ordered atomic environment. The e$_g^*$ and $t_{2g}^*$ features of the NS film, on the other hand, are broadened and less resolved. In comparison with the crystalline sample in the NS spectrum, the $t_{2g}^*$ energy position seems unaffected within our experimental resolution, while the e$_g^*$ are red-shifted of bout 0.4 eV. The overall effect is a transfer of spectral weight in the energy region between the $t_{2g}^*$ and e$_g^*$ bands. This is strong evidence of the distortion of the ligand octahedron \cite{Kasatikov2019, Ruzmetov2007} confirming the disordered nature of the NS samples. The e$_g^*$ orbitals i.e. the $d_{Z^2}$ and $d_{X^2-Y^2}$ orbitals, are the most affected by the geometric distortion. The red-shift in the XANES spectrum is a fingerprint of a reduced superposition between $V 3d - O 2p$ wave functions with the consequent destabilization of the e$_g^*$ bands. This spectral feature can be interpreted in terms of ligand geometric distortion as a symmetry breaking from $O_{h}$ to $D_{4h}$, with the bending of two opposite vertexes of the ligand octahedron \cite{Kasatikov2019} or with a combination of both. Since the e$_g^*$ orbitals are the most involved, we investigated V L edges line shape as a function of stoichiometry. In fact in vanadium oxides, the strong interaction between the 2p core hole and the 3d electrons in the final state is of the same order of magnitude of the spin-orbit splitting, causing a severe redistribution of the spectral weight of the entire spectrum toward high photon energy, i.e. toward the e$_g^*$ energy region, \cite{Zaanen1985, Abbate1991} making V L edges an ideal tool to study e$_g^*$ orbitals. \\

\begin{figure} [h!]
\includegraphics[width=0.5\textwidth]{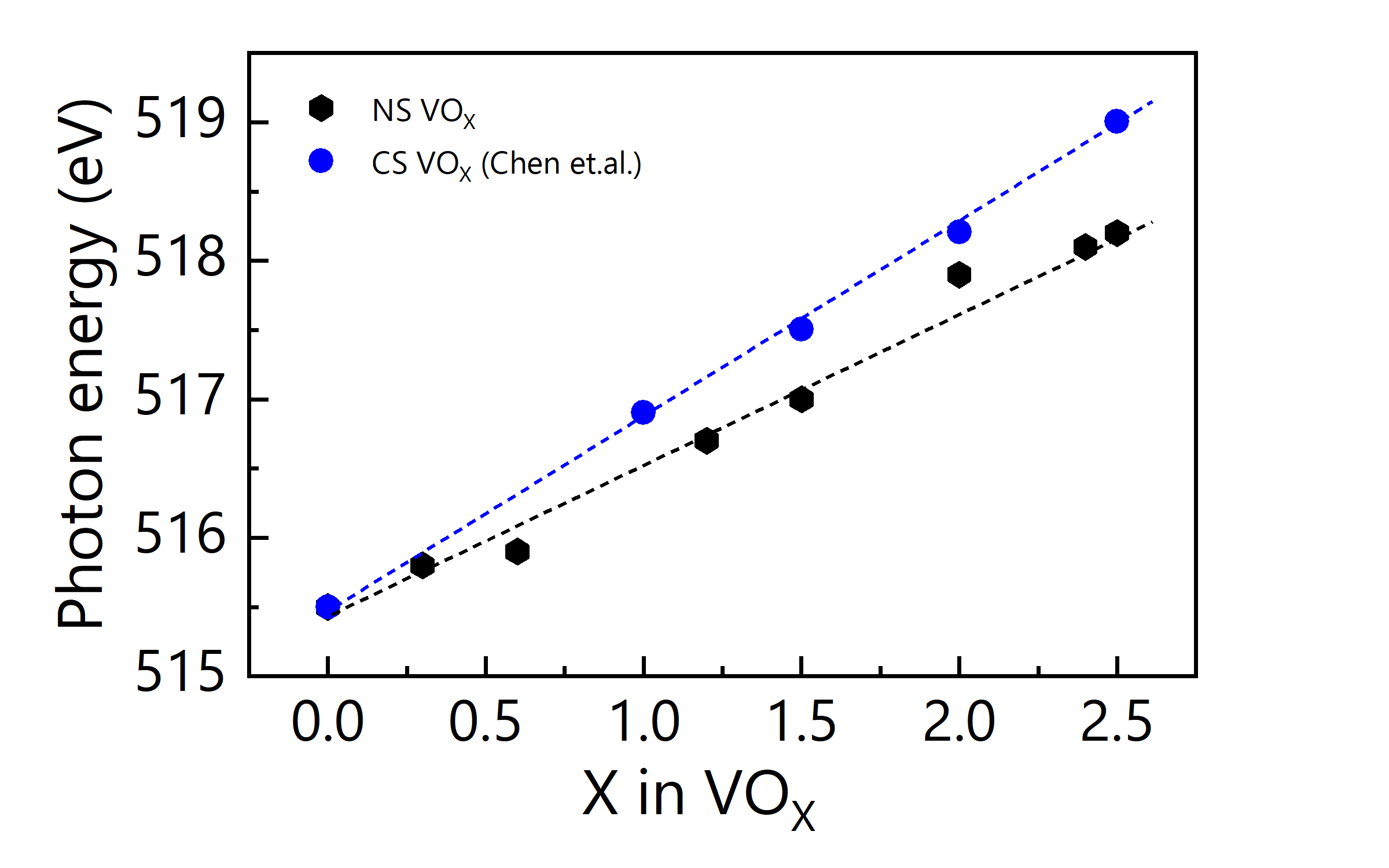}
\caption{Comparison between the V L$_3$ energy position as a function of stoichiometry for the NS VO$_x$ film (black) and crystalline samples data found in litterature (blue) \cite{Chen1994}.}
\label{rate}
\end{figure}

\begin{figure} [h!]
\includegraphics[width=0.7\textwidth]{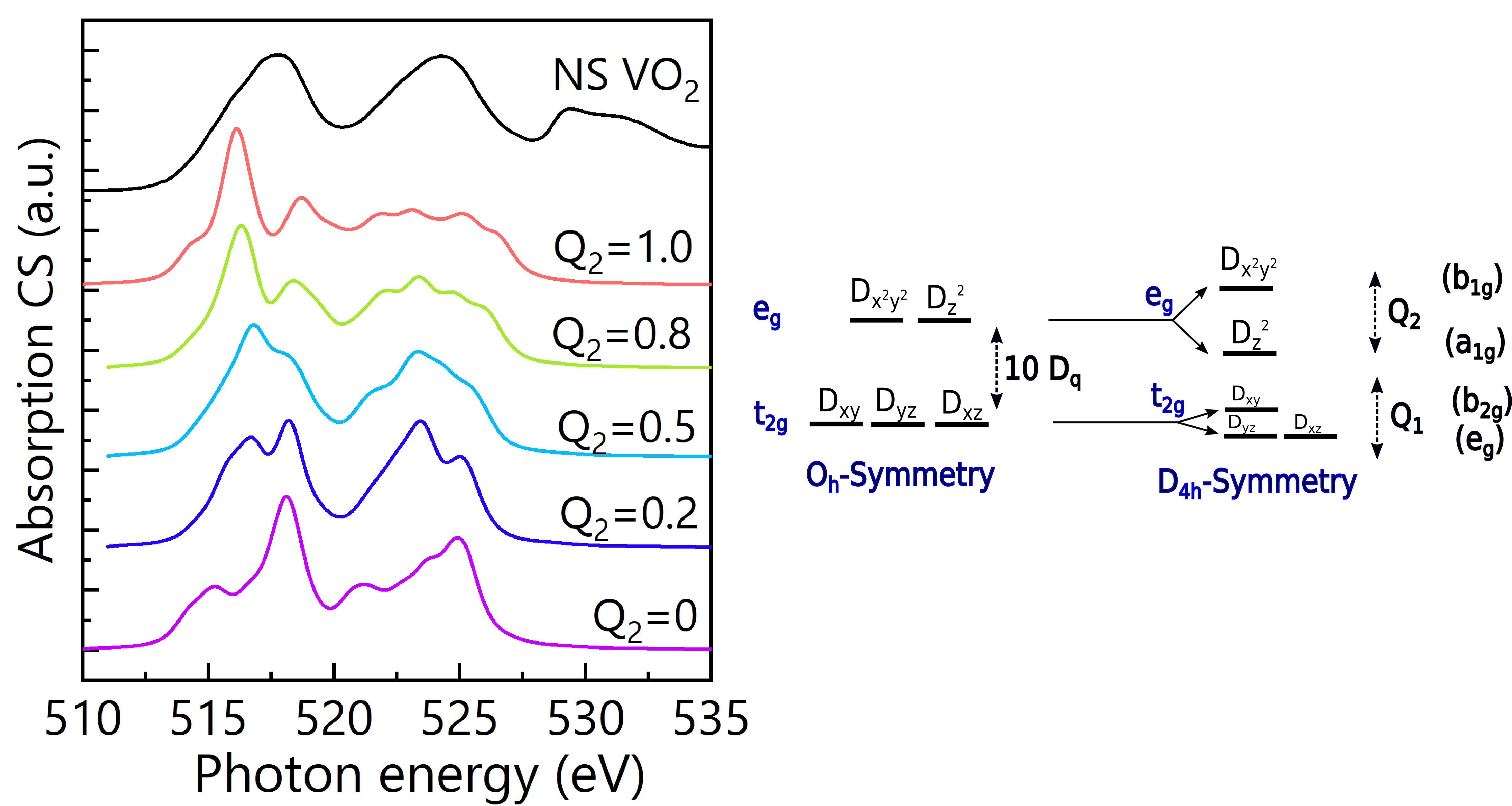}
\caption{Left: simulated V L edges spectra of  VO$_2$  for different Q$_2$ energy splitting values and NS VO$_2$ measured spectrum (black). Right: schematic representation of the energy levels splitting passing from $O_{h}$ to $D_{4h}$ symmetry.}
\label{sim}
\end{figure}

The vanadium L-edges spectrum shows two major components at $\sim 515.2$ eV (A) and $\sim 522.3$ eV (B) (see left panel of Figure \ref{VO}) in agreement with previously observed results \cite{Chen1997}. We could observe a chemical shift towards the higher photon energies with the introduction of the oxygen into the structure for values as low as $x=0.3$. The shift increases dramatically particularly after $x=1.5$ indicating the increase of the stoichiometry.
The L$_3$ maximum position evolution as a function of the oxidation state reveals a linearly increasing behavior by the increase of the oxygen concentration within the structure. The linear relationship to describe the stoichiometry dependence of L$_3$ and L$_2$ edges has been previously observed as well for bulk samples \cite{Chen1994}. However, a comparison of our data with that of reported in Ref\cite{Chen1994} evidence a relative L$_3$ redshift for the whole oxygen concentration range (except for pure vanadium where there is no oxygen in the structure, see Figure \ref{rate}). Furthermore, there are slight deviations from the linear behavior of the L$_3$ progress in our samples.
To establish the origin of the red-shifts and the deviations from linearity in our samples, several simulations were carried out within the charge transfer multiplet framework to investigate possible distortions effect on the VO$_x$ electronic structures. The calculations were performed by the configuration interaction cluster for octahedral and distorted VO$_2$ structures (see Figure \ref{sim}). The refinement of the analysis of the distorted structures was performed, employing the CTM4XAS \cite{StavitskiDegroot10} simulation code. A crystal field of 10Dq=2.35 eV was chosen as reported in Ref\cite{IkenoMizoguchi11}. Moreover, 80\% Slater integral reduction was performed and average Coulomb interaction $U_{dd}-U_{pd}$ equal to 0.88 eV was used based on the data in Ref\cite{IkenoMizoguchi11}. Overall full-width broadening was also chosen to be 0.35 eV. The evolution of the structure due to the distortion, by looking at the V L$_{2,3}$ edge, reflects the electronic occupation of the outer electronic shell by V $d$ electrons and the partial hybridization with the oxygen $p$ states. According to the diagram in Fig.\ref{sim}, in VO$_2$ the disorder leads to a tetragonal distortion. We applied the theoretical model to the V L$_{23}$ spectra taken from the NS VO$_2$ sample. The tetragonal distortion leads to an energy splitting of the $t_{2g}^*$ and e$_g^*$ orbitals resulting in the four levels e$_g$, b$_{2g}$, a$_{1g}$ and b$_{1g}$ (see Figure \ref{sim} right panel). The calculations were performed increasing the value of splitting Q$_2$= 0.2,0.5, 0.8 and 1.0 eV (splitting between a$_{1g}$ and b$_{1g}$) in order to simulate ligand displacement due to disorder and a constant Q$_1$ value of 0.2 eV of the t$_{2g}$ splitting . The choice to keep the Q$_1$ constant is justified considering the prominent sensitivity of V L edges to the e$_g^*$ orbitals which hide small spectral changes in the $t_{2g}^*$ energy region \cite{Zaanen1985}.
The theoretical spectra obtained for NS VO$_2$ and related distortion of the structure, by varying the values of the Q$_2$ (i.e., the e$_{g}$ splitting separation energy) are reported in Figure \ref{sim} (left panel) \cite{NohYeo06,StavitskiDegroot10}. These results suggest that the observed energy shift to lower photon energy values in our samples is mainly due to the presence of the distortion into the structure for which theoretical results with Q$_2$=0.5 eV correspond better with the experimental one. However, as it is visible in the figure the experimental spectra show features of distinct distorted states. Hence, it is reasonable to assume that the NS sample is characterized by the coexistence of different disordered local atomic geometries distorted respect to the ordered crystalline atomic arrangement. Our results suggest that the spectral features of NS vanadium oxides with distinct oxygen contents are results of the structural distortion induced by the small size of the nanoparticles. The finite dimensions of the NS films constituents and their irregular shape (see Figure \ref{tem}) provide a higher surface area respect to bulk counterpart, naturally leading to symmetry breaking and to creations of randomly distorted, i.e. disordered, atomic arrangements, which in last analysis influences the electronic structure of the sample. In addition, from Figure \ref{rate} it is observable how, respect to the bulk data, the V L$_{3}$ red-shift is stoichiometry dependent increasing as the oxygen content increases implying that the degree of disorder increases with stoichiometry. This can be understood considering that a high number of oxygen atoms corresponds to a high number of possible distorted atomic configurations achievable. As a consequence, the highest oxidation states are most likely to assume a severely distorted atomic arrangement respect to the samples with less oxygen.
This demonstrates that there is a complex interplay between samples dimensionality, stoichiometry, and disorder which can provide the possibility to modulate the electronic structure in VO$_x$ NS films to match the needs of advanced applications. 
   
\section{conclusions}

This research demonstrates the possibility to synthesize VO$_x$ films with controllable stoichiometry and disorder. The method allows to tune the stoichiometry of the NS film and its electronic structure respect to the bulk analogue. This capability can be exploited anytime electronic properties have to better match the application.
We have investigated the electronic structure evolution of several vanadium oxides NS films looking at the oxygen content inside the VO$_x$ matrix using in-situ XANES spectroscopy at the V L edges and O K edge. Moreover, the evolution of the $V 3d - O 2p$ hybridization was monitored from O K edge XANES spectra. We have demonstrated that broad $t_{2g}^*$ and e$_g^*$ features, as well as reduced $t_{2g}^*$ and e$_g^*$ splitting with respect to crystalline samples represent the fingerprint of locally disordered atomic structures.
Linear behavior of the V ${L_3}$ energy position has been observed as a function of the oxygen content. The result points a stoichiometry dependent redshift in relation to the bulk counterpart. 
We also showed that the transfer of spectral weight toward lower photon energy is induced by the local disorder, a direct effect of the dimension of the nanoparticles composing the film. Furthermore, the redshift increases along with the stoichiometry suggesting the NS VO$_x$ with high oxygen content growths with higher atomic disorder respect to the samples with lower oxygen content.
These stoichiometric features can be interpreted in terms of disordered ligands geometric arrangements. We showed the occurrence of a tetragonal distortion along with the bending of two opposite vertexes of the ligand octahedron that marks the coexistence of different distorted local environment within the NS films. 
Our study will be of great interest in hot-topic researches, such as Li-ion insertion applications, where the use of a disordered and NS material may be profitable. In addition, we provide a tool to monitor and control the disorder in NS VO$_x$ films which will be exploited in further studies on disordered materials.

\section{Acknowledgments}
The authors would like to thank D. Catone (ISM-CNR) for writing the acquisition software.
\bibliographystyle{unsrt}
\bibliography{manuscriptAD_V5}
\end{document}